\documentclass[12pt]{article}
\makeatletter
\@addtoreset{equation}{section}
\makeatother

\usepackage[pdftex]{graphicx}
\usepackage{subfigure}
\usepackage{array}
\usepackage{amsmath}

\usepackage{rotating}
\usepackage{longtable}
\usepackage{mathtools}
\usepackage{natbib}

\usepackage{array}
\usepackage{geometry}
\usepackage{color}
\usepackage{lipsum}

\usepackage{diagbox} 
\usepackage{pdfpages}
\usepackage{multicol}
\usepackage[utf8]{inputenc}
\usepackage[english]{babel}
\usepackage{mathtools}
\usepackage{bbm}
\usepackage{amssymb}
\usepackage{amsthm}
\usepackage{hyperref}
\usepackage{bm}
\usepackage{mathtools}
\usepackage{fancybox}
\usepackage{multirow}
\usepackage{caption}
\usepackage{subfigure}
\usepackage{algorithm}
\usepackage[affil-it]{authblk}
\usepackage[noend]{algpseudocode}
\usepackage{setspace}

\makeatletter
\newcommand*{\rom}[1]{\expandafter\@slowromancap\romannumeral #1@}
\newcommand{\tabincell}[2]{\begin{tabular}{@{}#1@{}}#2\end{tabular}}  
\makeatletter
\def\BState{\State\hskip-\ALG@thistlm}
\makeatother
\DeclarePairedDelimiter\abs{\lvert}{\rvert}%
\allowdisplaybreaks[4]

\usepackage{xcolor}
\renewcommand{\raggedright}{\leftskip=0pt \rightskip=0pt plus 0cm}
\title{{\bf Enhanced Structural Break Detection\\ in Functional Means}}
\author[]{Shuhao Jiao\thanks{Corresponding Author: shuhaojiao@cuhk.edu.hk}}
\author[]{Ngai Hang Chan}
\author[]{Chun Yip Yau}
\affil[]{Department of Statistics,\\The Chinese University of Hong Kong,\\ Shatin NT, Hong Kong}

\date{}
\begin{document}
	\maketitle			
	\setlength\parindent{0pt}
	\setlength{\parskip}{1em}
	\theoremstyle{definition}
	\newtheorem{theorem}{Theorem}
	\newtheorem{ass}{Assumption}
	\newtheorem{lemma}{Lemma}
	\newtheorem{remark}{Remark}
	\newtheorem{prop}{Proposition}
	\newtheorem{Definition}{Definition}
	\newtheorem{cor}{Corollary}
\begin{abstract}
A new change-point detector for structure breaks in functional means is developed in this paper. The detector is based on a novel easy-to-implement approach of dimension reduction. One major advantage of the proposed method is its efficiency in selecting the basis functions that capture the change/jump of functional means, leading to a higher detection power. 
We thoroughly investigate the asymptotic properties of the proposed detector when both the sample size and the incorporated dimension increase.  {The numerical simulation studies justify the superiority of the proposed approach compared to the existing competitors and highlight the necessity of aligning the basis functions with the change to be detected.} An application to annual humidity trajectories illustrates the practical superiority of the developed approach.\\

\noindent{\bf Key words}: Change point analysis, Change alignment, Dimension reduction, Functional Mean, Weakly dependent functional data.
\end{abstract}	
\section{Introduction}
This paper provides a new method to tackle a popular problem in functional data analysis, detecting the change point in functional means of a sequence of functional time series. The general setting is that a single change point partitions the entire sequence into two local stationary blocks, where the functions in each block share the same mean function. 

There have been a number of methods developed for functional structural breaks in mean function. Many of them are developed based on dimension reduction or projection. A typical step in projection-based approaches is to project the functions onto a finite number of basis functions, and the projection scores are employed to detect the change in mean of independent or dependent functional data sequence. See, for example, \cite{berkes2009detecting},  \cite{aue2009estimation}, \cite{zhang2011testing}, \cite{aston2012detecting} and the references therein. More recently, \cite{fremdt2014detection} consider structural break detection by using functional principal component analysis (fPCA) with an increasing number of projections. Dimension reduction is also utilized to detect change points of multivariate functions under separability assumptions (e.g., spatial temporal data or brain image data), see \cite{aston2012evaluating}, \cite{gromenko2017detection} and \cite{stoehr2021detecting}. Structural break detection in the coefficient operators of functional linear models is considered in \cite{aue2014dependent}. Structural break detection in spectrum and trace of covariance operator is studied in \cite{Jaruskova2013test} and \cite{aue2020structral}. Test for stationarity of functional time series in the spectral domain is developed in \cite{aue2020testing}. \cite{chiou2019identifying} and \cite{chiou2021greedy} study the multiple change-point problem for functional data.

In the change-point analysis of functional data, one major limitation of  dimension reduction is that, when the selected basis functions are not aligned with the jump of mean function, the projection-based detector fails to detect the change points. To solve this problem, an alternative fully functional approach is employed in \cite{horvath2014testing}, \cite{aue2018detecting} and \cite{jiao2022breaking}, which does not rely on dimension reduction. In the fully functional detection procedure, the null distribution involves infinitely many unknown parameters and requires additional truncation step, however. To circumvent this difficulty, \cite{Sharipov2016} and \cite{bucchia2017change} study the bootstrap procedure. In addition to the fully functional approach, \cite{torgovitski2015detecting} considered aligning the leading fPC with the change.

Although the fully functional detector is guaranteed to detect the change as the sample size increases, one major limitation of the approach is that it incorporates all basis functions that span the functional space, including potentially infinitely many irrelevant (unaligned with the change) basis functions. The irrelevant basis functions do not contribute to change point detection,  and can potentially lead to loss of detection power due to their nuisance effect. In the literature, sample trajectories are often pre-smoothed with a few smooth basis functions. In that case, the fully functional approach also performs decently for the smoothed functions, since the nuisance effect is significantly attenuated by functional smoothing, and thus is not very substantial when the number of pre-smoothing basis is kept small. Extended pre-smoothing can lead to a serious loss of information, however. It will make the fully functional detector fail when the functions are smoothed with low-frequency basis but the change functions are driven by high frequencies. Therefore, it is more advantageous to select basis functions which are informative to the change of mean, than to incorporate all basis functions or smooth the functions with pre-specified basis. 

In this paper, we develop a new detection method for structural breaks in functional means. The {\it key idea} is to align the selected basis functions with the change/jump function. To achieve this goal, we introduce a discrepancy enhanced covariance (DEC) operator, of which the eigenfunctions constitute the basis functions for dimension reduction. The DEC operator involves two parts. The first part is the long-run covariance and the second part is the enhancement term, which is calibrated to magnify the influence of the change-aligned basis functions. These basis functions have the advantage that they are aligned with the jump function. Unlike the fully functional approach, the null distribution of the developed detector only involves a finite number of parameters, and the nuisance effect of unaligned/irrelevant basis functions is substantially reduced. Another contribution of this paper is that we investigate the asymptotic properties under more complicated settings. Specifically, we allow both the change magnitude and the incorporated dimension to change with the sample size. We thoroughly investigate the regularity conditions under which the power of the proposed detector approaches one as the sample size goes to infinity, including the cases when the change magnitude diminishes.


The rest of the article is organized as follows. In Section \ref{s2}, we develop the change-aligned detection procedure and discuss the implementation details. Theoretical results are discussed in Section \ref{s3}. In Section \ref{s4}, we report the simulation results under different settings. In Section \ref{s5}, we present the real data analysis on annual humidity trajectories. The paper is concluded in Section \ref{s6}. Proofs are given in the online supplementary material.

\section{Change-aligned Detection Procedure}
\label{s2}
\subsection{Projection-based Detector}
A single change-point model can be formulated as
\begin{equation}
\label{model}
X_n(t)=\left\{
\begin{aligned}
\mu_0(t)+e_n(t),\qquad n\le k^*,\\
\mu_1(t)+e_n(t),\qquad n>k^*,
\end{aligned}
\right.
\end{equation}
where $k^*=\lfloor N\theta^*\rfloor$ and $\theta^*$ is the scaled location of the change-point in [0,1], and the zero-mean random functions $\{e_n(t)\colon n\in\mathbb{N}\}$ take realizations in $L^2[0,1]$ and satisfy that $E\int e^2_n(t)dt<\infty$. In the space $L^2[0,1]$, the inner product of two elements $x(t),y(t)$ are defined as $\langle x,y\rangle=\int_0^1x(t)y(t)dt$ and the norm is defined as $\|x\|=\{\int_0^1x^2(t)dt\}^{1/2}<\infty$. It is assumed that $\{e_n(t)\colon n\in\mathbb{N}\}$ are weakly dependent as quantified in the following assumption.
\begin{ass}
\label{lp}
There is a measurable function $f\colon S^\infty\to L^2[0,1]$, where $S$ is a measurable space, and the $i.i.d.$~innovations $\{\epsilon_n\colon n\in\mathbb{N}\}$ take values in $S$, so that $e_n(t)=f(\epsilon_n,\epsilon_{n-1},\ldots)$. In addition, there exists a $m$-dependent sequence $\{e_{n,m}(t)\colon i\in\mathbb{N}\}$, so that $e_{n,m}(t)=f(\epsilon_n,\ldots,\epsilon_{n-m+1},\epsilon^*_{n-m},\epsilon^*_{n-m-1},\ldots),$
where $\epsilon_n^*$ is an independent copy of $\epsilon_n$, such that
$\sum_{m=0}^\infty\{E\|e_n(t)-e_{n,m}(t)\|^{p}\}^{1/p}<\infty$ for some $p>2$.
\end{ass}

{A process is termed $L^p$-m approximable if it satisfies Assumption \ref{lp} (see \cite{hormann2010weakly}). This is a mild assumption which is satisfied by many processes, such as auto-regressive processes and moving average processes.}

In this paper, a single structural break problem is considered. 
When there are multiple change-points, we propose to apply some localization method to segment the whole sequence into multiple blocks, where at most one change-point (AMOC) assumption is made for each block, and then use the proposed approach to each block to detect the change-point. More details can be found in the real data analysis of humidity trajectories. This is beyond the scope of this paper, and we do not pursue its details here.

Therein, the goal is to detect whether a change point exists and to identify the location of the change-point. Define the jump function as $\delta(t)=\mu_0(t)-\mu_1(t)$, and the following test is implemented to detect the change point,
\begin{equation}
\label{test}
H_0:\delta(t)=0\ \mbox{for almost all}\ t  \ \ \mbox{vs}\ \ H_a:\delta(t)\ne0\ \mbox{for a non-negligible set of}\ t.
\end{equation}
Given a sequence of basis functions $\{b_d(t)\colon d\ge1\}$, suppose that $X_n(t)=\sum_{d\ge1}\eta_{nd}b_d(t)$ and let $\bm{\eta}_{n}=(\eta_{nd},\ldots,\eta_{nD})'$ where $D>0$, then the cumulative summation (CUSUM) is defined as 
\begin{align*}
\bm{S}_{N,\theta}=\sum_{n=1}^{\lfloor N\theta\rfloor}\bm{\eta}_n-\theta\sum_{n=1}^N\bm{\eta}_n.
\end{align*} 
The projection-based method is based on the squared CUSUM statistics,
\begin{equation}
\label{proj-cusum}
T_N(\theta)=N^{-1}\|\bm{S}_{N,\theta}\|_2^2,
\end{equation}
where $\|\cdot\|_2$ denotes the $\ell^2$-norm. 
The value of $T_N(\theta)$ should be large at the true change point $\theta^*$, thus by convention, the following max-type quantity is employed as the detector of change point
$$T_N(\hat{\theta}_N^*)=\max_{0<\theta<1}T_N(\theta),$$
and for uniqueness, the infimum of the maximizers of $T_N(\theta)$, namely $$\hat{\theta}_N^*=\inf\{\theta:T_N(\theta)= \sup_{\theta'\in(0,1)} T_N(\theta')\}$$ is assumed to be the change point candidate. In principle, it is desirable for the selected basis functions $\{b_d(t)\colon d=1,\ldots,D\}$ to capture the jump function, say, $\langle b_d,\delta\rangle\ne0$ for some $d$. Otherwise, the method fails even if $\|\delta\|$ is much bigger than zero.

The following result quantifies the null distribution of the projection-based detector.
\begin{theorem} 
\label{th1}
Under Assumption \ref{lp} and $H_0$, $$T_N(\hat{\theta}_N^*)\overset{d}\to\sup_{\theta\in(0,1)}\bm{B}'(\theta)\Sigma_D\bm{B}(\theta),$$ where $\bm{B}(\theta)=(B_1(\theta),\ldots,B_D(\theta))'$ and $\{B_d(\theta)\colon d\ge1\}$ are $i.i.d.$~Brownian bridges and $\Sigma_D=\sum\limits_{h=-\infty}^{\infty}\mbox{Cov}(\bm{\eta}_{n},\bm{\eta}_{n+h})$. 
\end{theorem}
The theorem follows from Theorem A.1 in \cite{aue2009dependent}.
Theorem \ref{th1} asymptotically validates the test of $H_0$. Specifically, $H_0$ is rejected if the test statistic $T_N(\hat{\theta}_N^*)$ exceeds the corresponding quantile of the null distribution $\sup\limits_{\theta\in(0,1)}\bm{B}'(\theta)\Sigma_D\bm{B}(\theta)$. 

\subsection{Selection of Basis Functions}
\label{sofbf}
Define the (auto)covariance function of $\{X_n\colon n\in\mathbb{N}\}$ as $C_{X,h}(t,s)=E\{e_n(t)e_{n+h}(s)\}$, and the long-run covariance as $LC_X(t,s)=\sum_{h=-\infty}^\infty C_{X,h}(t,s)$. Assume that, with a sequence of positive and decreasing eigenvalues $\{\tau_d\colon d\ge1\}$ and orthonormal eigenfunctions $\{\phi_d(t)\colon d\ge1\}$, the spectral decomposition $LC_X(t,s)=\sum_{d\ge1}\tau_d\phi_d(t)\phi_d(s)$ is allowed.

The selection of $\{b_d(t)\colon d=1,\ldots,D\}$ highly influences the performance of the detector. 
A popular way of selecting the basis functions is to employ the major eigenfunctions of the (long-run) covariance operator $\mathcal{LC}_X(\cdot)$, induced by the kernel $LC_X(t,s)$, and the resulting null distribution is $\sup_{\theta\in(0,1)}\sum_{d=1}^D\tau_dB^2_d(\theta)$ (see \cite{berkes2009detecting}, \cite{hormann2015dynamic} and \cite{torgovitski2015detecting}). Such basis functions are not guaranteed to align with $\delta(t)$. To solve this problem, our  approach is based on the major eigenfunctions of the discrepancy enhanced covariance (DEC) operator described as follows. 

To separate the jump-aligned component and other irrelevant components, 
first transform the functions as follows:
\begin{align}
\label{trans}
Y^{(\kappa)}_n(t)=X_n(t)-\left\langle X_n,\frac{\delta}{\|\delta\|+\kappa}\right\rangle\frac{\delta(t)}{\|\delta\|+\kappa},
\end{align} 
where $\kappa$ is a small-valued positive tuning parameter shrinking to zero as $N\to\infty$. Note that $\delta(t)$ is typically unknown, and the estimation of $\delta(t)$ will be discussed in Section \ref{est}. 

The term $Y_n^{(\kappa)}(t)$ in (\ref{trans}) is well defined under both $H_0$ and $H_a$. The DEC is then defined as 
\begin{align*}
K^{(\kappa)}(t,s)=LC_{Y,\kappa}(t,s)+\rho\delta(t)\delta(s),
\end{align*}
where $\rho$ is the enhancement parameter to be specified. The quantity $LC_{Y,\kappa}(t,s)$ is the long-run covariance of $Y_n^{(\kappa)}(t)$, defined as
\begin{align*}
LC_{Y,\kappa}(t,s)=\sum_{h=-\infty}^\infty C_{Y,h}^{(\kappa)}(t,s),
\end{align*} 
where $C_{Y,h}^{(\kappa)}(t,s)=\mbox{cov}\{Y^{(\kappa)}_n(t),Y^{(\kappa)}_{n+h}(s)\}=E\{Y^{(\kappa)}_n(t)-E{Y^{(\kappa)}_n(t)}\}\{Y^{(\kappa)}_{n+h}(s)-EY^{(\kappa)}_{n+h}(s)\}$. 
By Mercer's theorem, suppose that a sequence of decreasing positive eigenvalues $\{\theta^{(\kappa)}_d\colon d\ge1\}$ and a sequence of corresponding orthonormal eigenfunctions $\{\psi^{(\kappa)}_d\colon d\ge1\}$ can be found such that 
$$K^{(\kappa)}(t,s)=\sum_{d\ge1}\theta^{(\kappa)}_d\psi^{(\kappa)}_d(t)\psi^{(\kappa)}_d(s).$$
It is proposed to make use of $\{\psi_d^{(\kappa)}(t)\colon d=1,\ldots,D\}$ in defining the test statistic $T_N(\theta)$. 

To understand this selection procedure, first define $K(t,s)=LC_Y(t,s)+\rho\delta(t)\delta(s)$ under $H_a$, where $$LC_Y(t,s)=\sum_{h=-\infty}^\infty C_{Y,h}(t,s)=\sum_{h=-\infty}^\infty E\{Y_n(t)-E{Y_n(t)}\}\{Y_{n+h}(s)-EY_{n+h}(s)\}$$ and $\{Y_n(t)\colon n\in\mathbb{N}\}$ are defined as
\begin{equation}
\label{yn}
Y_n(t)=X_n(t)-\left\langle X_n,\frac{{\delta}}{\|{\delta}\|}\right\rangle\frac{{\delta}(t)}{\|{\delta}\|}.
\end{equation}
The kernel function $K(t,s)$ is positive definite, thus similar to $K^{(\kappa)}(t,s)$, the spectral decomposition can be found as $K(t,s)=\sum_{d\ge1}\theta_d\psi_d(t)\psi_d(s).$

Evidently, under $H_a$ and as $\kappa\to0$, $LC_{Y,\kappa}\to LC_{Y}$, and therefore,  $\{(\theta_d^{(\kappa)},\psi_d^{(\kappa)}(t))\colon d\ge1\}$ converge to $\{(\theta_d,\psi_d(t))\colon d\ge1\}$. Clearly, $\delta(t)$ is orthogonal to all the eigenfunctions of $LC_Y(\cdot)$ due to the projection (\ref{yn}), and thus $\{\rho\|\delta\|^2,\delta(t)/\|\delta\|\}$ is a pair of eigenvalue and eigenfunction of $K(t,s)$.
In what follows, it is supposed that $\theta_{d^*}=\rho\|\delta\|^2$ and $\psi_{d^*}(t)=\delta(t)/\|\delta\|$, and denote $\theta^{(\kappa)}_{d^*}$ and $\psi^{(\kappa)}_{d^*}(t)$ as the counterparts of $K^{(\kappa)}(t,s)$. Observe that $\psi_{d^*}(t)=\delta(t)/\|\delta\|$ is the only eigenfunction of $K(t,s)$ aligned with the jump function $\delta(t)$, and a large value of $\rho$ leads to large eigenvalue $\rho\|\delta\|^2$. 

In practice, $\delta(t)$ is typically unknown and $K(t,s)$ may be not well-defined (when $\|\delta\|=0$ under $H_0$, and this leads to the non-consistency of $\hat{\delta}/\|\hat{\delta}\|$). To solve this problem, we add the tuning parameter $\kappa$ and obtain $\{Y^{(\kappa)}_n(t),n\ge1\}$. It is of major interest to enhance the influence of the jump-aligned counterpart of $K^{(\kappa)}(t,s)$, namely, $\psi_{d^*}^{(\kappa)}(t)$. 
In the following, $\psi_{d^*}^{(\kappa)}(t)$ is termed {\it jump-aligned basis function}.

The step (\ref{trans}) is important for the selection of dimension $D$ (see Section \ref{sofd}).
An alternative approach is to employ the major functional principal components of $LC_X(t,s)+\rho\delta(t)\delta(s)$. While this is also reasonable, we still recommend to do the projection (\ref{trans}) first. The reason is that, 
without projection, it is hard to find all aligned (not orthogonal to $\delta(t)$) eigenfunctions of $LC_X(t,s)+\rho\delta(t)\delta(s)$, making the selection of $D$ much more complicated.
{
}
\subsection{Estimations}
\label{est}
Since the detector depends on the first $D$ (the selection of $D$ will be discussed in Section \ref{sofd}) eigenfunctions of $K^{(\kappa)}(t,s)$, one key step is to estimate $K^{(\kappa)}(t,s)$. {First, we discuss the estimation of $\delta(t)$. We propose to segment the entire functional sequence at the midpoint into two separate subsequences, $\{X_n(t)\colon n=1,\ldots,\lfloor N/2\rfloor\}$ and $\{X_n(t)\colon n=\lfloor N/2\rfloor+1,\ldots,\lfloor N\rfloor\}$, and estimate $\hat{\delta}(t)$ as follows
\begin{align*}
\hat{\delta}(t)=\frac{1}{\lfloor N/2\rfloor}\sum_{n=1}^{\lfloor N/2\rfloor}X_n(t)-\frac{1}{N-\lfloor N/2\rfloor+1}\sum_{n=\lfloor N/2\rfloor+1}^{N}X_n(t).
\end{align*}}

Observe that, under $H_a$, $E\{\hat{\delta}(t)\}/\delta(t)=C(\theta^*)\le1$, where $C(\theta^*)$ is a constant related to $\theta^*$, and the equality holds when $k^*=\lfloor N/2\rfloor$. Although $\hat{\delta}(t)$ is not a consistent estimator of $\delta(t)$, this is not an issue in our approach. The goal here is to find the basis functions that are aligned with the jump function. In other words, the shape rather than the magnitude of $\delta(t)$ is of the major interest. 
Thus the non-consistency of estimation does not lead to any loss of information. 

Then we construct $Y_n^{(\kappa)}(t)$ as
\begin{align*}
Y^{(\kappa)}_n(t)=X_n(t)-\left\langle X_n,\frac{\hat{\delta}}{\|\hat{\delta}\|+\kappa}\right\rangle\frac{\hat{\delta}(t)}{\|\hat{\delta}\|+\kappa}.
\end{align*} 
Further this leads to the empirical (auto)covariance of $\{Y_n^{(\kappa)}(t)\colon n\in\mathbb{N}\}$ displayed below
\begin{align*}
\widehat{C}^{(\kappa)}_{Y,h}(t,s)&=\frac{1}{N-h}\sum_{n=1}^{N-h}\{Y^{(\kappa)}_n(t)-\bar{Y}^{(\kappa)}_n(t)\}\{Y^{(\kappa)}_{n+h}(s)-\bar{Y}^{(\kappa)}_{n+h}(s)\},\qquad h\ge0,\\
\widehat{C}^{(\kappa)}_{Y,h}(t,s)&=\frac{1}{N+h}\sum_{n=|h|+1}^{N}\{Y^{(\kappa)}_n(t)-\bar{Y}^{(\kappa)}_n(t)\}\{Y^{(\kappa)}_{n+h}(s)-\bar{Y}^{(\kappa)}_{n+h}(s)\},\qquad h<0,
\end{align*}
where 
\begin{equation*}
\bar{Y}^{(\kappa)}_n(t)=\left\{
\begin{aligned}
&\frac{1}{\hat{k}_N^{(f)}}\sum_{j=1}^{\hat{k}_N^{(f)}}{Y}^{(\kappa)}_j(t),\qquad 1\le n\le\hat{k}_N^{(f)},\\
&\frac{1}{N-\hat{k}_N^{(f)}}\sum_{j=\hat{k}_N^{(f)}+1}^{N}{Y}^{(\kappa)}_j(t),\qquad \hat{k}_N^{(f)}+1\le n\le N,
\end{aligned}
\right.
\end{equation*}
and $\hat{k}_N^{(f)}$ is defined as the infimum of the the maximizer(s) of the following quantity
\begin{align*}
M(k)=\frac{1}{N}\int\left(\sum_{n=1}^kX_n(t)-\frac{k}{N}\sum_{n=1}^NX_n(t)\right)^2dt.
\end{align*}
The estimation of $K^{(\kappa)}(t,s)$ is then given as follows,
\begin{align*}
\widehat{K}^{(\kappa)}(t,s)&=\sum_{h=-\ell}^\ell W\left(\frac{h}{\ell}\right)\widehat{C}^{(\kappa)}_{Y,h}(t,s)+\rho\hat{\delta}(t)\hat{\delta}(s),
\end{align*}
where $W(\cdot)$ is the kernel function, and $\ell$ is the bandwidth.
 See, e.g., \cite{rice2017gregory} for the selection of the bandwidth $\ell$. 

The corresponding empirical eigenfunctions are defined through the eigen-equation $\widehat{K}^{(\kappa)}(\hat{\psi}_d^{(\kappa)})(t)=\hat{\theta}_d^{(\kappa)}\hat{\psi}_d^{(\kappa)}(t)$, leading to $\hat{\bm{\eta}}_n=(\hat{\eta}_{n1},\ldots,\hat{\eta}_{nD})'$, where $\hat{\eta}_{nd}=\langle X_n,\hat{\psi}_d^{(\kappa)}\rangle$ and $\widehat{\bm{S}}_{N,\theta}=\sum_{n=1}^{\lfloor N\theta\rfloor}\hat{\bm{\eta}}_{d}-\theta\sum_{n=1}^N\hat{\bm{\eta}}_{d}$. Similarly, $\Sigma_D$ is estimated with the kernel estimator 
\begin{align*}
\widehat{\Sigma}_D=\sum_{h=-\ell}^\ell W\left(\frac{h}{\ell}\right)\widehat{C}_{\eta,h},
\end{align*}  
where $\widehat{C}_{\eta,h}$ is defined similar to $\widehat{C}^{(\kappa)}_{Y,h}$ with $Y_n^{(\kappa)}(t)$ replaced by $\hat{\bm{\eta}}_n$. 
The estimated detector is obtained as 
$\widetilde{T}_N(\theta)=N^{-1}\|\widehat{\bm{S}}_{N,\theta}\|^2_2$. 

\subsection{Selection of $\rho$ and $\kappa$}
\label{s3.1}
Note that we only need to do the enhancement when $H_a$ is true, and the enhancement term only brings more estimation uncertainty under $H_0$. 
Therefore, it is ideal that the enhancement term lays asymptotically trivial influence under $H_0$. To achieve this goal, $\rho$ should be selected so that $\rho\hat{\delta}(t)\hat{\delta}(s)$ converges to zero faster than $\widehat{LC}_{Y,\kappa}(t,s)-E\{LC_{Y,\kappa}(t,s)\}$, say, $$E\|\widehat{LC}_{Y,\kappa}(t,s)-E\{\widehat{LC}_{Y,\kappa}(t,s)\}\|^2/\{\rho^2E\|\hat{\delta}(t)\|^4\}\to\infty.$$ Assume that $E\|\hat{\delta}\|^4/(E\|\hat{\delta}\|^2)^2<\infty,$  then equivalently
\begin{equation*}
E\|\widehat{LC}_{Y,\kappa}(t,s)-E\{\widehat{LC}_{Y,\kappa}(t,s)\}\|^2/\{\rho E\|\hat{\delta}(t)\|^2\}^2\to\infty.
\end{equation*}
Under Assumption \ref{lp}, it can be deduced that $NE\|\widehat{LC}_{Y,\kappa}(t,s)-E\{\widehat{LC}_{Y,\kappa}(t,s)\}\|^2<\infty$ for any fixed $\kappa$ and $\ell$ (see Theorem 4.1 in \cite{hormann2010weakly} and Lemma 4 in \cite{jiao2022breaking}), and it can also be deduced that $NE\|\hat{\delta}(t)\|^2<\infty$ under Assumption \ref{lp} (see Lemma 2 in \cite{jiao2022breaking}). Therefore, we propose to choose $\rho=N^{\beta},$
with some $0<\beta<1/2$.

For the identifiability of the jump aligned basis function, we propose to adjust the selected $\rho$ so that $\rho\|\hat{\delta}\|^2$ lies in the middle of the two neighboring eigenvalues of $\widehat{\mathcal{LC}}_{Y,\kappa}(\cdot)$. Suppose that 
$$\widehat{LC}_{Y,\kappa}(t,s)=\sum_{d\ge1}\hat{\lambda}_d^{(\kappa)}\hat{\nu}^{(\kappa)}_d(t)\hat{\nu}^{(\kappa)}_d(s),$$
where $\hat{\lambda}_1^{(\kappa)}>\hat{\lambda}_2^{(\kappa)}>\cdots$. If $\rho\|\hat{\delta}\|^2$ is greater than the maximal eigenvalue of $\widehat{\mathcal{LC}}_{Y,\kappa}(\cdot)$, then $\rho$ is selected so that $\rho\|\hat{\delta}\|-\hat{\lambda}^{(\kappa)}_1$ is greater than a non-trivial positive value $L_\rho$, e.g., $\hat{\lambda}^{(\kappa)}_1-\hat{\lambda}^{(\kappa)}_2$. The identifiability of $\psi_{d^*}^{(\kappa)}(t)$ will be discussed in Section \ref{ie}. 

The principle of selecting $\kappa$ is that the term $\hat{\delta}/(\|\hat{\delta}\|+\kappa)$ in Eq.\ (\ref{trans}) converges to zero in probability as $N\to\infty$ under $H_0$ to solve the non-consistency problem ($\hat{\delta}/\|\hat{\delta}\|$ is not consistent and not well-defined under $H_0$), or equivalently $\kappa^{-2} E\|\hat{\delta}\|^2\to0$ under $H_0$. Therefore, we propose that $\kappa=N^{-\alpha_\kappa}\{\int\widehat{LC}_X(t,t)\,dt\}^{1/2}$ with some $0<\alpha_\kappa<1/2$. We require $\alpha_\kappa>0$, since otherwise $\kappa$ would mitigate the enhancement. Here, the role of  $\int\widehat{LC}_X(t,t)\,dt$ is to attenuate the effect of data variation.
In the simulation section, it is shown that the detection performance is {\it robust} to the selection of $\rho$ and $\kappa$. 


\subsection{Selection of $D$}
\label{sofd}
Cumulative percentage of variance is a widely accepted criterion for the selection of dimension, and is adjusted for the selection of $D$ in our new method. 
Here, the goal is to incorporate the jump-aligned basis function ${\psi}^{(\kappa)}_{d^*}(t)$, of which the corresponding eigenvalue converges to $\rho\|{\delta}\|^2$ under $H_a$ as $\kappa\to0$. In principle, $D$ should be selected so that 1) the selected basis functions explain sufficient data variation to attenuate the nuisance effect of the estimation deviation $\hat{\delta}(t)-E\hat{\delta}(t)$ and control the type-I error, and 2) $D>d^*$ to incorporate the jump-aligned basis function. More details are now discussed.

Define
\begin{equation} 
\label{rule}
R_\lambda(D)=\sum_{d=1}^D\hat{\lambda}^{(\kappa)}_d\bigg/\sum_{d\ge1}\hat{\lambda}^{(\kappa)}_d,
\end{equation}
$\gamma$ as a positive constant taking the value, e.g., 80\%--90\%, $D_{pre}$ as the minimal value of $D$ satisfying $R_\lambda(D)\ge\gamma$, and $\hat{d}^*$ as the minimal value of $d$ satisfying $\hat{\lambda}_d^{(\kappa)}<\rho\|\hat{\delta}\|^2$. 
Two scenarios are considered:
\begin{itemize}
\item[(1)]if $\rho\|\hat{\delta}\|^2>\hat{\lambda}^{(\kappa)}_{D_{pre}}$, set $D> D_{pre}$,
\item[(2)]if $\rho\|\hat{\delta}\|^2\le\hat{\lambda}^{(\kappa)}_{D_{pre}}$, set $D>\hat{d}^*.$ 
\end{itemize}

Now we give the reasoning of the selection. 
Condition (1) is important in controlling the size of the test. 
As an extreme case, if $\hat{\psi}_{d^*}^{(\kappa)}$ is the only incorporated basis function, the type-I error can be much higher than the nominal level, since the estimation errors falsely favor a change-point under $H_0$. To attenuate such ``over-enhancement" effect, it is necessary to incorporate multiple basis functions that capture sufficient variation to mitigate the estimation uncertainty of $\hat{\delta}(t)$. Condition (1) and (2)\ substantially increases the chance that the jump-aligned basis $\psi_{d^*}^{(\kappa)}$ is selected, 
 and is important in solving the ``non-alignment" problem in projection-based detector. 

\section{Theoretical Results}
\label{s3}
\subsection{Convergence rate $\widehat{\mathcal{LC}}_{Y,\kappa}(t,s)$}
In this section, we present the convergence rate of $\widehat{\mathcal{LC}}_{Y,\kappa}(t,s)$ under both $H_0$ and $H_a$.
In what follows, the quantity ``const." represents a positive constant and $\|\cdot\|_\mathcal{S}$ signifies the Hilbert-Schmidt norm. First we introduce some notations.
\begin{ass}
\label{auto}
There exist $\alpha_c,\alpha_\kappa>0$, so that $\|C_{X,h}\|_2\le \mbox{const.}\,h^{-\alpha_c}$ and $\kappa=O(N^{-\alpha_\kappa})$. 
\end{ass}
\begin{ass}
\label{kernel}
$c^{-1}_1|u|^{\alpha_w} \le 1-W(u)\le c_1|u|^{\alpha_w}$ for $|u|\le1$ and some $c_1\ge0$, $W(0)=1$, $0\le W(\cdot)\le1$, $W(u)=W(-u)$, $W(u)=0$ if $|u|>1$, and the bandwidth $\ell$ satisfies $\ell=O(N^{\alpha_\ell})$, where $0<\alpha_\ell<1/2$. 
\end{ass}
{Assumption \ref{kernel} is suitable for a general class of kernel functions $W(u)$. When $\alpha_w=1$, the triangular kernel
\begin{equation*}
W(u)=\left\{
\begin{aligned}
&1-|u|,\qquad |u|\le1,\\
&0,\qquad |u|>1,
\end{aligned}
\right.
\end{equation*}
satisfies the assumption. When $\alpha_w=0$, uniform kernel 
\begin{equation*}
W(u)=\left\{
\begin{aligned}
&1,\qquad |u|\le1,\\
&0,\qquad |u|>1,
\end{aligned}
\right.
\end{equation*}
satisfies the assumptions. Other values of $\alpha_w$ indicate polynomial decay rates of $W(u)$.}

The following theorem quantifies the convergence rate of the estimated long-run covariance operator $\widehat{\mathcal{LC}}_{Y,\kappa}(\cdot)$ induced by $\widehat{LC}_{Y,\kappa}(t,s)$.
\begin{theorem}
\label{th6}
Under Assumption \ref{lp}, \ref{auto}, and \ref{kernel}, if $H_0$ is true and $N^{-1}\ell\kappa^{-2}\to0$, then for an arbitrarily small $\epsilon>0$,
\begin{align*}
\|\widehat{\mathcal{LC}}_{Y,\kappa}-\mathcal{LC}_{X}\|_\mathcal{S}&\le O_p(1)N^{\max\{\alpha_\ell-1/2,-(\alpha_c-1)/\alpha_\ell,-1+2\alpha_\kappa+\alpha_\ell\}}\\
&\vee\left\{
\begin{array}{rcl}
N^{-(\alpha_c-1)\alpha_\ell}, & & \mbox{if}\ {\alpha_w-\alpha_c>-1}.\\
N^{-(\alpha_c-1)\alpha_\ell+\epsilon},& & \mbox{if}\ {\alpha_w-\alpha_c=-1}.\\
N^{-\alpha_w\alpha_\ell}, & & \mbox{if}\ {\alpha_w-\alpha_c<-1}.
\end{array} \right.
\end{align*}
Moreover, if $H_a$ is true and $\ell\kappa\|\delta\|^{-1}\to0$,  then for an arbitrary small $\epsilon>0$,
\begin{align*}
\|\widehat{\mathcal{LC}}_{Y,\kappa}-\mathcal{LC}_{Y}\|_\mathcal{S}&\le O_p(1)N^{\max\{\alpha_\ell-1/2,-(\alpha_c-1)/\alpha_\ell,\alpha_\ell-\alpha_\kappa-\alpha_\delta\}}\\
&\vee\left\{
\begin{array}{rcl}
N^{-(\alpha_c-1)\alpha_\ell}, & & \mbox{if}\ {\alpha_w-\alpha_c>-1}.\\
N^{-(\alpha_c-1)\alpha_\ell+\epsilon},& & \mbox{if}\ {\alpha_w-\alpha_c=-1}.\\
N^{-\alpha_w\alpha_\ell}, & & \mbox{if}\ {\alpha_w-\alpha_c<-1}.
\end{array} \right.
\end{align*}
\end{theorem}
Theorem \ref{th6} gives the convergence rate of $\widehat{\mathcal{LC}}_{Y,\kappa}$ under both $H_0$ and $H_a$. The convergence rate is specified through multiple parameters.  To simplify notations, $r_0$ is denoted as the convergence rate of $\widehat{\mathcal{LC}}_{Y,\kappa}$ under $H_0$, and $r_a$ is denoted as the convergence rate of $\widehat{\mathcal{LC}}_{Y,\kappa}$ under $H_a$. 

Denote $T^o_N(\theta)$ to be the projection-based test statistics based on the eigenfunctions of $\mathcal{LC}_X(\cdot)$. We develop the following theorem.
\begin{theorem} 
\label{uniformcon}
Under Assumption \ref{lp}, \ref{auto}, \ref{kernel}, and $H_0$, if $D\to\infty$ and $N^{-r_0}\sum_{d=1}^D\delta^{-1}_{\tau,d}\to0$, then uniformly for $\theta\in(0,1)$, 
$\widetilde{T}_N(\theta)\overset{d}\to T^o_N(\theta)$, where $\delta_{\tau,1}=\tau_1-\tau_2$, $\delta_{\tau,d}=\max\{\tau_{d-1}-\tau_d,\tau_{d}-\tau_{d+1}\}$ for $d\ge2$. 
\end{theorem}

The theorem illustrates that, under some regularity conditions, the null distribution of the new change-aligned detector is asymptotically equivalent to that of the ordinary fPC-based detector. Consequently, $\widetilde{T}_N(\hat{\theta}_N^*)\overset{d}\to\sup_{\theta\in(0,1)}\sum_{d=1}^D\tau_dB^2_d(\theta)$.

\subsection{Selection of $\psi_{d^*}^{(\kappa)}$ and Identifiability}
\label{ie}
In this section, we investigate the performance of change-alignment and the identifiability of the basis functions 
selected by the approach in Section \ref{sofd}.
Under $H_a$, the non-zero enhancement term $\rho\delta(t)\delta(s)$ is tuned through $\rho$, and the identifiability of $K^{(\kappa)}(t,s)$ might be violated if $\rho$ is not selected judiciously. 
Since $LC_{Y,\kappa}(t,s)$ and $LC_{Y}(t,s)$ are positive definite, by Mercer's theorem, we can find a sequence of decreasing positive values and orthonormal basis functions for each of them so that,
\begin{align*}
LC_{Y,\kappa}(t,s)=\sum_{d\ge1}{\lambda}^{(\kappa)}_d\nu^{(\kappa)}_d(t)\nu^{(\kappa)}_d(s), \qquad LC_{Y}(t,s)=\sum_{d\ge1}{\lambda}_d\nu_d(t)\nu_d(s).
\end{align*}
Since $LC_{Y,\kappa}(t,s)\to LC_Y(t,s)$ under $H_a$ as $\kappa\to0$, the identifiability of $\{\psi^{(\kappa)}_d(t)\colon d\ne d^*\}$ is asymptotically guaranteed  given the identifiability of $\{\nu_d(t)\colon d\ge1\}$. Thus the identifiability of $\psi^{(\kappa)}_{d^*}(t)$ is of major interest here. 
{Recall that} the selected $\rho$ is adjusted so that $\rho\|\hat{\delta}\|^2$ lies in the middle of the two neighboring eigenvalues of $\widehat{\mathcal{LC}}_{Y,\kappa}(\cdot)$, and if $\rho\|\hat{\delta}\|^2$ is greater than the maximal eigenvalue of $\widehat{\mathcal{LC}}_{Y,\kappa}(\cdot)$, then $\rho$ is selected so that $\rho\|\hat{\delta}\|-\hat{\lambda}^{(\kappa)}_1$ is greater than a positive value $L_\rho$. It can be shown that the identifiability of $\psi_{d^*}^{(\kappa)}(t)$ is asymptotically guaranteed under some mild conditions. To justify this, we first introduce the following assumptions.
\begin{ass}
\label{tuning}
Under $H_a$, the eigenvalues $\{\lambda_d\colon d\ge1\}$ satisfy the conditions $R^{-1}d^{-\alpha_\theta}\le\lambda_d\le Rd^{-\alpha_\theta}$, where $R$ is a positive constant, and $\lambda_d-\lambda_{d+1}\ge\mbox{const.} d^{-\alpha_\theta-1}$. 
\end{ass}
Assumption \ref{tuning} quantifies the decay rate of eigenvalues $\{\lambda_d\colon d\ge1\}$ and restricts the spacing of eigenvalues from being overly small, which enables the identifiability of $\{\lambda_d\colon d\ge1\}$ (see also \cite{cai2006prediction}). 
We define $\Delta_{\rho,D}=\abs{\hat{\lambda}_{d^*-1}-\hat{\lambda}^{(\kappa)}_{D-1}}$, which depends on the selection of $D$ and $\rho$.
\begin{theorem}
\label{iden}
Under Assumption \ref{lp}--\ref{tuning}, and $H_a$, if $D^{\alpha_\theta+1}N^{-r_a}\to0$ 
and $\Delta_{\rho,D}^{-1}\ell\kappa\|\delta\|^{-1}\to0$, then asymptotically almost surely, the following events are true:\\
1) $D>d^*$,\\ 
2) For an arbitrary $\epsilon>0$, $\max\{\hat{\theta}^{(\kappa)}_{d^*-1}-\hat{\theta}^{(\kappa)}_{d^*},\hat{\theta}^{(\kappa)}_{d^*}-\hat{\theta}^{(\kappa)}_{d^*+1}\}\ge (\theta_{d^*-1}-\theta_{d^*+1})/(2+\epsilon)$ or $\hat{\theta}^{(\kappa)}_{d^*}-\hat{\theta}^{(\kappa)}_2\ge L_\rho/(1+\epsilon)$ as $d^*=1$.
\end{theorem}

\begin{remark}
See the definition of $\{\theta_d\colon d\ge1\}$ and $\{\theta^{(\kappa)}_d\colon d\ge1\}$ in Section \ref{sofbf}.
\end{remark}

Theorem \ref{iden} demonstrates that 
the identifiability of the jump-aligned basis is guaranteed and the basis functions selected by the adjusted variance portion criterion contain the jump-aligned basis as $N\to\infty$. 

\subsection{Power Studies}
This section presents the asymptotic properties and the detection power of the new detector. It is known that, when 1) the number of projections and the jump function is fixed, and 2) some of the selected basis functions are aligned with the jump function, the power of CUSUM-type detector approaches one as the sample size goes to infinity (see \cite{berkes2009detecting}). However, it has not been studied under what conditions the asymptotically perfectly performed detector (with power approaching one) can be achieved in a more general setting. In contrast, the dimension $D$, the magnitude of jump $\|\delta\|$, and the tuning parameters $\rho$ and $\kappa$ are all allowed to vary with the sample size $N$ here. 
To present the theoretical results under this general setting, we first introduce the function
\begin{equation*}
V(\theta)=\left\{
\begin{aligned}
\theta(1-\theta^*),&\qquad 0<\theta\le \theta^*,\\
\theta^*(1-\theta),&\qquad \theta^*<\theta<1,
\end{aligned}
\right.
\end{equation*}
and the following assumption.
\begin{ass}
\label{a5}
$\rho= O(N^{\beta})$, and $\|\delta\|=O(N^{\alpha_\delta})$, where $\alpha_\delta>-1/2$.
\end{ass}
The assumption $\alpha_\delta>-1/2$ restricts that the change magnitude shrinks to zero no faster than the estimation error.
The following theorem presents the convergence of $\widetilde{T}_N({\theta})$.
\begin{theorem}
\label{convergence}
Under Assumptions \ref{lp}--\ref{a5} and $H_a$, define $r_\delta=\max\{\alpha_\ell-\alpha_\kappa-\alpha_\delta-1/2,\alpha_\delta+\beta-1/2\}$ and 
$U_N=\max\{D^{1/2}(\rho\|\delta\|^2)^{-1},N^{r_\delta-r_a}(\rho\|\delta\|^2)^{-(1+1/\alpha_\theta)},N^{r_\delta-1/2}\},$ if $D^{\alpha_\delta+3/2}/N^{r_a}\to0$ and $D(\rho\|\delta\|^2)^{1/\alpha_\theta}\to\infty$, then
\begin{equation*}
\sup_{\theta\in(0,1)}\left|N^{-1}\widetilde{T}_N(\theta)-\|\delta\|^2V^2(\theta)\right|\le O_p(N^{-r_\delta})\|\delta\|U_N.
\end{equation*}
\end{theorem}
Since it is assumed that $\alpha_\delta>-1/2$, $N\|\delta\|^2\to\infty$ as $N\to\infty$. Note that $d^*=O((\rho\|\delta\|^2)^{-1/\alpha_\theta})$ under Assumption \ref{tuning}, and the condition $D(\rho\|\delta\|^2)^{1/\alpha_\theta}\to\infty$ ensures that the jump-aligned basis function is selected.
The theorem demonstrates that the convergence rate of $N^{-1}\widetilde{T}_N(\theta)$ is uniformly bounded by $N^{-r_\delta}\|\delta\|U_N$. Therefore, if the ratio $N^{-r_\delta}\|\delta\|U_N/\|\delta\|^2$ converges to zero, then it is sufficient to conclude that $\widetilde{T}_N(\theta)\overset{p}\to N\|\delta\|^2V^2(\theta)$ uniformly for $\theta\in(0,1)$, which leads to $\widetilde{T}_N(\theta)\overset{p}\to\infty$.

Based on Theorem \ref{convergence}, the following corollary presents the regularity conditions that guarantees the detection power approaching one.
\begin{cor}
\label{corollary}
Under Assumptions \ref{lp}--\ref{tuning} and $H_a$, if the conditions in Theorem  \ref{convergence} hold, and $N^{-r_\delta}\|\delta\|^{-1}U_N\to0$, then $\mbox{Pr}(H_0\mbox{ is rejected}|H_a)\to1$. 
In addition, $\hat{\theta}_N\overset{p}\to\theta^*$.
\end{cor}
The first part of the corollary can be obtained from Theorem \ref{convergence}, say, $\widetilde{T}_N(\theta)\overset{p}\to\infty$ under $H_a$. 
The consistency of $\hat{\theta}_N$ can be obtained by the continuous mapping theorem of ``argmax" function and the fact that $\theta^*$ is the unique maximizer of $V(\theta)$.

\section{Simulation}
\label{s4}
\subsection{General Setting}
\label{setting}
Finite sample properties are investigated in this section. First, $N=200$ or $N=400$ identically distributed functions are generated over the unit interval $[0,1]$ with $B=20$ Fourier basis functions $\{F_6(t),\ldots,F_{25}(t)\}$ specified as follows
\begin{equation*}
F_i(t)=\left\{
\begin{array}{ccl}
1, & & \text{if}\ i=1.\\
\sqrt{{2}}\cos(2\pi {k}t), & & \text{if}\ i=2k.\\
\sqrt{{2}}\sin(2\pi {k}t), & &  \text{if}\ i=2k+1.
\end{array} \right. 
\end{equation*}
The change in functional means is located in the middle of the sequence and is driven by the 2nd Fourier basis, which is orthogonal to the $20$ basis functions $\{F_6(t),\ldots,F_{25}(t)\}$. The final functions are simulated by the following basis expansion contaminated by random noises $\{\epsilon_n(t_j)\colon n=1,\ldots,N,\ j=1,\ldots,T\}$
\begin{align*}
X_n(t_j)=\left\{
\begin{aligned}
\sum_{d=1}^{B}\xi_{nd}F_{d+5}(t_{j})+\epsilon_n(t_j),&\qquad 1\le n\le \lfloor N/2\rfloor,\ j=1,\ldots,T\\
\sum_{d=1}^{B}\xi_{nd}F_{d+5}(t_j)+\delta(t_j)+\epsilon_n(t_j),&\qquad \lfloor N/2\rfloor+1\le n\le N,\ j=1,\ldots,T,
\end{aligned}
\right.
\end{align*}
where $\epsilon_n(t_j)\overset{i.i.d.}\thicksim\mathcal{N}(0,s^2)$, and $t_j=j/100$, $T=100$.
We set $\delta(t)=aF_{2}(t)$, where $a=0$ under $H_0$ and $a>0$ under $H_a$. To highlight the effect of the magnitude $\|\delta\|$, different values of $a$ are considered. The variation of random errors $\{\epsilon_n(t_j)\colon t_j=1,\ldots,T, n\ge1\}$ is tuned through $s$. The obtained functions are smoothed with the first 35 Fourier basis functions. 

Define $\bm{\sigma}=\mbox{diag}(1.2^{-2},1.2^{-4},\ldots,1.2^{-2B})$. Two distributional setups of $\bm{\xi}_n=(\xi_{n1},\ldots,\xi_{nB})$ are considered, namely, $\{\bm{\xi}_n\colon n\ge1\}$ are
\begin{itemize}
\item[1.] (Independent case) $i.i.d.$~random vectors following the distribution $\mathcal{N}(0,\bm{\sigma})$.
\item[2.] (Dependent case) a FMA(3) process $\bm{\xi}_n=\sum_{j=1}^3\Phi_j\bm{e}_{n-j}+\bm{e}_n,$ where $\bm{e}_n\sim\mathcal{N}(0,\bm{\sigma})$, $\Phi_1=0.6\bm{I}_B$, $\Phi_2=0.4\bm{I}_B$, $\Phi_3=0.2\bm{I}_B$ and $\bm{I}_B$ is the $B\times B$ identity matrix.
\end{itemize}
We use the R package {\it sde} to simulate the null distribution, and the simulations are run with the same seed under different settings. The proposed detector (denoted by CA) is compared with two other competitors, which are representative in change-point detection, say, the fPC-based approach (denoted by fPC, see e.g., \cite{berkes2009detecting}) and the fully functional approach (denoted by FF, see e.g., \cite{aue2018detecting}). For the fPC-based detector, the dimension is selected so that the incorporated functional principal components explain 90\% of the total data variation. 

\subsection{Empirical Size and Power}
\label{power}
In this section, we compare the empirical size and power of the four methods. 
In each setting, the simulation runs are repeated for 5000 times at nominal level $0.05$. Here we set $\gamma=90\%$. The enhancement parameters $\rho$ considered are $\rho_1=N^{0.25}$, $\rho_2=N^{0.3}$, $\rho_3=N^{0.35}$, and $\rho_4=N^{0.40}$. The tuning parameter $\kappa$ considered here is $\kappa=\kappa_1=N^{-0.4}\{\int\widehat{LC}_X(t,t)\,dt\}^{1/2}$. 
The empirical sizes and powers are reported in Table \ref{t1} for setting 1 and Table \ref{t2} for setting 2. 
We also test the performance of the new detector when $\kappa=\kappa_2=N^{-0.35}\{\int\widehat{LC}_X(t,t)\,dt\}^{1/2}$, $\kappa=\kappa_3=N^{-0.3}\{\int\widehat{LC}_X(t,t)\,dt\}^{1/2}$ under the $i.i.d.$\ setting to show its robustness to $\kappa$, see Table \ref{t3}.

\begin{table}[H]
\centering
\captionsetup{justification=centering}

	\caption{Empirical sizes and powers under different values of $a$ and $s$ (i.i.d., $\kappa=\kappa_1$).}
	\begin{tabular}{|c|c|c|p{0.4in}p{0.4in}p{0.4in}p{0.4in}|p{0.4in}p{0.4in}|}
	\hline 
         $a$  & $N$ & $s$ &\multicolumn{4}{c|}{\tabincell{c}{CA\\$\rho_1$\hspace{0.9cm} $\rho_2$\hspace{0.9cm}  $\rho_3$\hspace{0.9cm}  $\rho_4$}}  & FF & fPC\\
         \hline 
         \multirow{6}{*}{0.00}&\multirow{3}{*}{200}				&0.5&0.056&0.056&0.057&0.058&0.056&0.048\\
         &&1.0&0.057&0.057&0.057&0.057&0.057&0.046\\
         &&1.5&0.056&0.057&0.057&0.057&0.052&0.047\\
         \cline{2-9}
         	&\multirow{3}{*}{400}&0.5&0.047&0.048&0.049&0.048&0.053&0.049\\
	 &&1.0&0.055&0.054&0.054&0.056&0.050&0.049\\
	 &&1.5&0.058&0.057&0.058&0.058&0.050&0.044\\
	 \hline 
	 
	 \multirow{6}{*}{0.20}&\multirow{3}{*}{200}&0.5&0.148&0.149&0.148&0.148&0.145&0.055\\
         &&1.0&0.139&0.139&0.139&0.141&0.126&0.073\\
         &&1.5&0.143&0.143&0.144&0.144&0.126&0.103\\
         \cline{2-9}
         	&\multirow{3}{*}{400}&0.5&0.410&0.406&0.400&0.401&0.363&0.056\\
	 &&1.0&0.403&0.410&0.420&0.426&0.360&0.141\\
	 &&1.5&0.444&0.447&0.448&0.444&0.373&0.313\\
	 \hline 
	 
	 \multirow{6}{*}{0.22}&\multirow{3}{*}{200}&0.5&0.183&0.186&0.186&0.185&0.179&0.057\\
         &&1.0&0.176&0.176&0.178&0.178&0.158&0.099\\
         &&1.5&0.184&0.183&0.186&0.185&0.159&0.134\\
         \cline{2-9}
         	&\multirow{3}{*}{400}&0.5&0.605&0.599&0.595&0.598&0.544&0.058\\
	 &&1.0&0.597&0.604&0.614&0.623&0.537&0.289\\
	 &&1.5&0.627&0.633&0.629&0.629&0.550&0.481\\
	 \hline 
	 
	 \multirow{6}{*}{0.24}&\multirow{3}{*}{200}&0.5&0.241&0.240&0.238&0.241&0.232&0.060\\
         &&1.0&0.232&0.232&0.234&0.234&0.207&0.144\\
         &&1.5&0.238&0.238&0.239&0.237&0.209&0.183\\
         \cline{2-9}
         	&\multirow{3}{*}{400}&0.5&0.844&0.841&0.833&0.836&0.791&0.061\\
	 &&1.0&0.816&0.823&0.837&0.838&0.775&0.603\\
	 &&1.5&0.816&0.818&0.818&0.814&0.761&0.704\\
	 \hline 
	\end{tabular}
	\label{t1}
\end{table}

\begin{table}[H]
\centering
\captionsetup{justification=centering}

	\caption{Empirical sizes and powers under different values of $a$ and $s$ (FMA, $\kappa=\kappa_1$).}
	\begin{tabular}{|c|c|c|p{0.4in}p{0.4in}p{0.4in}p{0.4in}|p{0.4in}p{0.4in}|}
	\hline 
         $a$  & $N$ & $s$ &\multicolumn{4}{c|}{\tabincell{c}{CA\\$\rho_1$\hspace{0.9cm} $\rho_2$\hspace{0.9cm}  $\rho_3$\hspace{0.9cm}  $\rho_4$}}  & FF & fPC\\
         \hline 
         \multirow{6}{*}{0.00}&\multirow{3}{*}{200}&2.0&0.054&0.055&0.054&0.054&0.053&0.022\\
         &&3.0&0.050&0.050&0.051&0.051&0.050&0.021\\
         &&4.0&0.051&0.050&0.051&0.051&0.047&0.022\\
         \cline{2-9}
         	&\multirow{3}{*}{400}&2.0&0.051&0.052&0.053&0.053&0.057&0.037\\
	 &&3.0&0.053&0.054&0.055&0.054&0.051&0.035\\
	 &&4.0&0.049&0.050&0.049&0.050&0.047&0.032\\
	 \hline
	 
	 \multirow{6}{*}{0.40}&\multirow{3}{*}{200}&2.0&0.123&0.121&0.120&0.121&0.114&0.056\\
         &&3.0&0.115&0.115&0.115&0.117&0.108&0.054\\
         &&4.0&0.111&0.112&0.111&0.112&0.099&0.051\\
         \cline{2-9}
         	&\multirow{3}{*}{400}&2.0&0.280&0.284&0.291&0.287&0.287&0.225\\
	 &&3.0&0.307&0.306&0.303&0.300&0.267&0.222\\
	 &&4.0&0.288&0.286&0.283&0.284&0.258&0.202\\
	 \hline 
	 
	 \multirow{6}{*}{0.50}&\multirow{3}{*}{200}&2.0&0.203&0.200&0.198&0.198&0.184&0.104\\
         &&3.0&0.190&0.191&0.190&0.188&0.175&0.095\\
         &&4.0&0.182&0.181&0.182&0.183&0.162&0.092\\
         \cline{2-9}
         	&\multirow{3}{*}{400}&2.0&0.655&0.667&0.676&0.672&0.667&0.592\\
	 &&3.0&0.692&0.693&0.679&0.688&0.614&0.566\\
	 &&4.0&0.641&0.643&0.631&0.621&0.582&0.507\\
	 \hline 
	 
	 \multirow{6}{*}{0.60}&\multirow{3}{*}{200}&2.0&0.361&0.363&0.362&0.367&0.328&0.222\\
         &&3.0&0.345&0.343&0.342&0.343&0.313&0.201\\
         &&4.0&0.331&0.334&0.333&0.331&0.291&0.189\\
         \cline{2-9}
         	&\multirow{3}{*}{400}&2.0&0.983&0.984&0.984&0.986&0.985&0.973\\
	 &&3.0&0.977&0.978&0.976&0.979&0.964&0.959\\
	 &&4.0&0.958&0.958&0.955&0.945&0.939&0.910\\
	 \hline

	\end{tabular}
	\label{t2}
\end{table}

\begin{table}[H]
\centering
\captionsetup{justification=centering}

	\caption{Empirical sizes and powers of the new method (i.i.d., $\kappa=\kappa_2$ and $\kappa_3$).}
	\begin{tabular}{|c|c|c|p{0.4in}p{0.4in}p{0.4in}p{0.4in}|p{0.4in}p{0.4in}p{0.4in}p{0.4in}|}
	\hline 
         $a$  & $N$ & $s$ &\multicolumn{4}{c|}{\tabincell{c}{CA ($\kappa=\kappa_2$)\\$\rho_2$\hspace{0.9cm} $\rho_2$\hspace{0.9cm}  $\rho_3$\hspace{0.9cm}  $\rho_4$}}  & \multicolumn{4}{c|}{\tabincell{c}{CA ($\kappa=\kappa_3$)\\$\rho_1$\hspace{0.9cm} $\rho_2$\hspace{0.9cm}  $\rho_3$\hspace{0.9cm}  $\rho_4$}}  \\
         \hline 
         \multirow{6}{*}{0.00}&\multirow{3}{*}{200}				&0.5&0.056&0.057&0.058&0.058&0.056&0.056&0.058&0.058\\
         &&1.0&0.055&0.056&0.055&0.056&0.055&0.055&0.056&0.056\\
         &&1.5&0.056&0.056&0.057&0.057&0.056&0.057&0.057&0.057\\
         \cline{2-11}
         	&\multirow{3}{*}{400}&0.5&0.047&0.048&0.049&0.049&0.048&0.048&0.047&0.047\\
	 &&1.0&0.054&0.053&0.054&0.055&0.051&0.052&0.053&0.053\\
	 &&1.5&0.057&0.057&0.058&0.058&0.057&0.056&0.057&0.057\\ 
	 \hline 
	 
	 \multirow{6}{*}{0.20}&\multirow{3}{*}{200}&0.5&0.149&0.148&0.148&0.147&0.149&0.149&0.149&0.149\\
         &&1.0&0.139&0.139&0.139&0.139&0.140&0.139&0.141&0.139\\
         &&1.5&0.144&0.144&0.143&0.145&0.143&0.144&0.143&0.144\\         
         \cline{2-11}
         	&\multirow{3}{*}{400}&0.5&0.411&0.405&0.400&0.403&0.409&0.407&0.401&0.403\\
	 &&1.0&0.402&0.410&0.419&0.427&0.405&0.408&0.417&0.425\\
	 &&1.5&0.445&0.448&0.447&0.443&0.448&0.445&0.446&0.444\\
	 \hline 
	 
	 \multirow{6}{*}{0.22}&\multirow{3}{*}{200}&0.5&0.184&0.184&0.185&0.186&0.188&0.185&0.186&0.185\\
         &&1.0&0.173&0.175&0.177&0.177&0.175&0.175&0.178&0.176\\
         &&1.5&0.184&0.183&0.185&0.185&0.184&0.182&0.185&0.185\\
         \cline{2-11}
         	&\multirow{3}{*}{400}&0.5&0.604&0.603&0.592&0.596&0.605&0.600&0.591&0.598\\
	 &&1.0&0.596&0.604&0.614&0.621&0.600&0.605&0.614&0.621\\
	 &&1.5&0.630&0.633&0.633&0.631&0.632&0.633&0.632&0.633\\
	 \hline 
	 
	 \multirow{6}{*}{0.24}&\multirow{3}{*}{200}&0.5&0.242&0.239&0.239&0.240&0.240&0.241&0.239&0.241\\
         &&1.0&0.232&0.233&0.234&0.234&0.232&0.233&0.233&0.235\\
         &&1.5&0.237&0.238&0.238&0.239&0.236&0.238&0.238&0.238\\       
         \cline{2-11}
         	&\multirow{3}{*}{400}&0.5&0.844&0.839&0.833&0.836&0.845&0.839&0.834&0.837\\
	 &&1.0&0.818&0.822&0.836&0.838&0.816&0.823&0.833&0.839\\
	 &&1.5&0.817&0.820&0.821&0.816&0.819&0.820&0.823&0.818\\
	 \hline 
	\end{tabular}
	\label{t3}
\end{table}
\subsection{Variation of the Detected Change-points}
\begin{figure}[H]
\center
\includegraphics[scale=0.55]{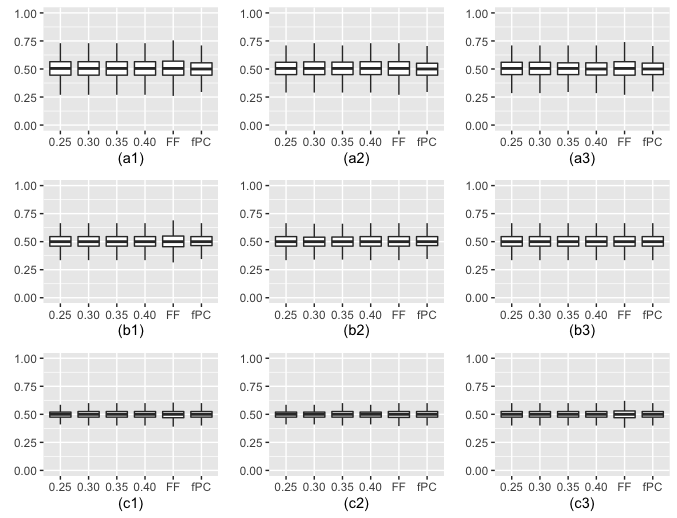}
\caption{Box-plots of the detected change-points (N=200, FMA).}
\label{ex1}
\end{figure}

\begin{figure}[H]
\center
\includegraphics[scale=0.55]{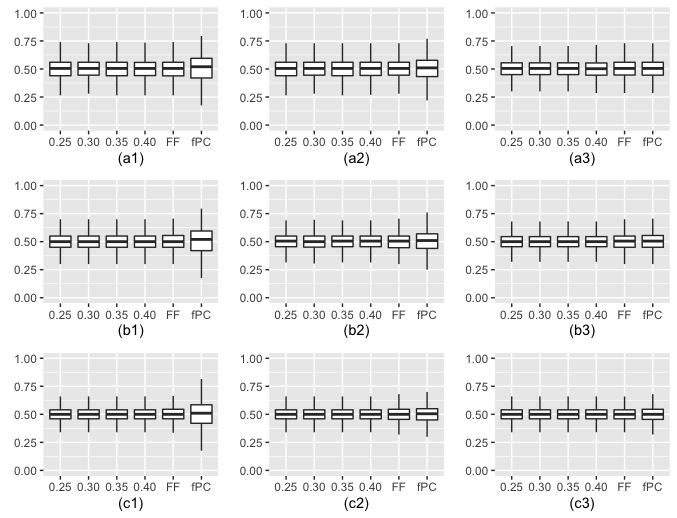}
\caption{Box-plots of the detected change-points (N=200, i.i.d.).}
\label{ex2}
\end{figure}

\begin{figure}[H]
\center
\includegraphics[scale=0.55]{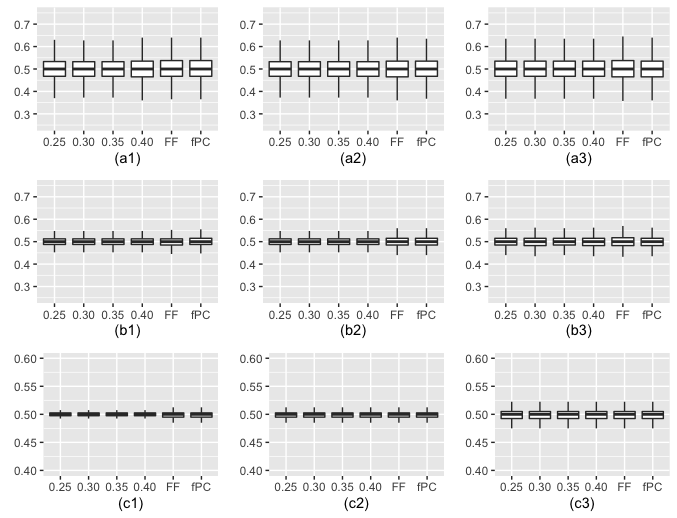}
\caption{Box-plots of the detected change-points (N400, FMA).}
\label{ex3}
\end{figure}

\begin{figure}[H]
\center
\includegraphics[scale=0.55]{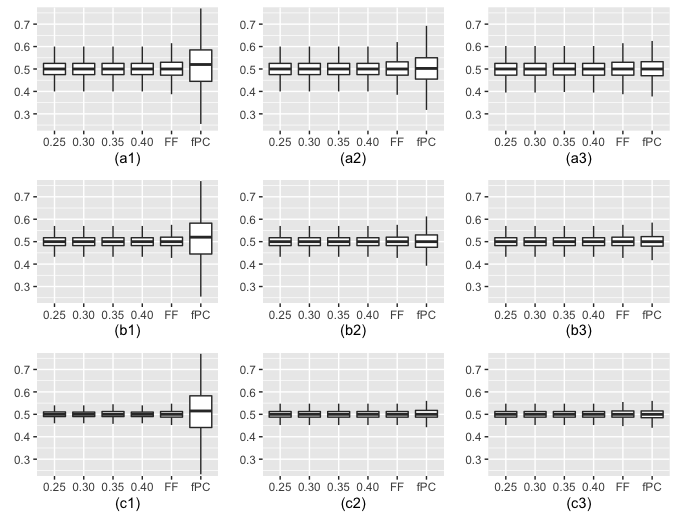}
\caption{Box-plots of the detected change-points (N=400, i.i.d.).}
\label{ex4}
\end{figure}

To study the variation of the detected change-points, we provide the box-plots of the detected change-points in Figures \ref{ex1}--\ref{ex4}. In each figure, there are six boxes. The first four boxes pertains to the proposed detector under $\rho=N^{0.25}$, $\rho=N^{0.3}$, $\rho=N^{0.35}$, and $\rho=N^{0.4}$ respectively. The 5th box pertains to the fully functional detector and the last one pertains to the fPC-based approach. Overall, the variance of the detected change-points of the proposed detector and the fully functional detector are similar, and that of the fPC-based procedure can be sometimes much higher. 

In Figure \ref{ex1} and \ref{ex3}, subfigures (a1)--(a3) correspond to the cases of $a=0.4$, subfigures (b1)--(b3) correspond to the cases of $a=0.5$, and subfigures (c1)--(c3) correspond to the cases of $a=0.6$. The first column corresponds to the cases of $s=2$, the second column corresponds to the cases of $s=3$, and the third column corresponds to the cases of $s=4$.

In Figure \ref{ex2} and \ref{ex4}, subfigures (a1)--(a3) correspond to the cases of $a=0.2$, subfigures (b1)--(b3) correspond to the cases of $a=0.22$, and subfigures (c1)--(c3) correspond to the cases of $a=0.24$. The first column corresponds to the cases of $s=0.5$, the second column corresponds to the cases of $s=1$, and the third column corresponds to the cases of $s=1.5$.
\vspace{2cm}

\subsection{Necessity of Change Alignment}
\begin{figure}[H]
\center
\includegraphics[scale=0.62]{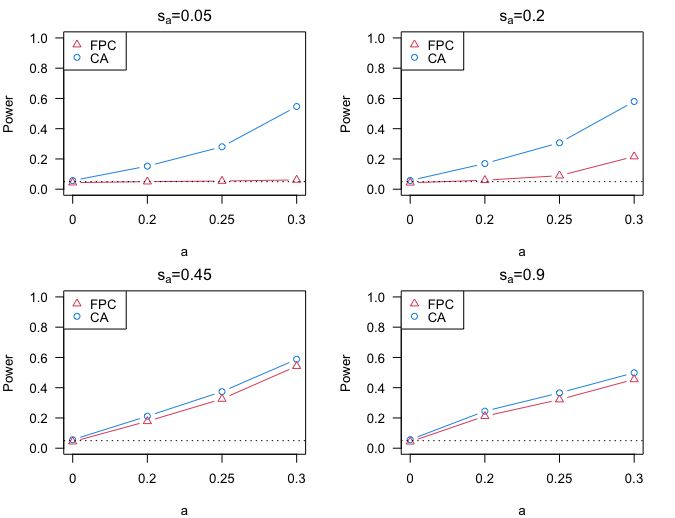}
\caption{Power Comparison. The dotted black line signifies the significance level.}
\label{powercom}
\end{figure}

{To thoroughly investigate the necessity of aligning the basis functions with the jump function, we examine more comparisons between the new approach and the fPC-based approach. We consider a variety of cases where the alignment between the major eigenfunction of $LC_X(t,s)$ and the jump function $\delta(t)$ changes. Specifically, in addition to the 20 basis functions used to generate $\{X_n(t)\colon n\ge1\}$, the 2nd Fourier basis, which drives the jump function, is also incorporated in the simulation here. Specifically, $\{X_n(t)\colon n\ge1\}$ can be expressed as
\begin{align*}
X_n(t)=\left\{
\begin{aligned}
\xi_n^aF_2(t)+\sum_{d=1}^{20}\xi_{nd}F_{d+5}(t),&\qquad 1\le n\le \lfloor N/2\rfloor,\\
\xi_n^aF_2(t)+\sum_{d=1}^{20}\xi_{nd}F_{d+5}(t)+\delta(t),&\qquad \lfloor N/2\rfloor+1\le n\le N,
\end{aligned}
\right.
\end{align*}
see Section \ref{setting} for the details of $\delta(t)$. Observe that, 
only $F_2(t)$ is aligned with the potential mean change.
Here, $\xi_n^a\overset{i.i.d.}\sim\mathcal{N}(0,s_a^2)$, $s_a=0.9, 0.45, 0.2, 0.05$, and $\bm{\xi}_n$ are simulated under the independent case as described in Section \ref{setting}. The role of $s_a$ is to tune the alignment between the eigenfunctions of $LC_X(t,s)$ and the jump function, and a large value of $s_a$ leads to a high rank of the jump-aligned function in the set of eigenfunctions of $LC_X(t,s)$, which makes it easier to select the jump-aligned basis for the fPC-based approach.

Here, we set $\rho=N^{0.4}$, and $N=200$. The sizes/powers of the two approaches are displayed in Figure \ref{powercom}. From the results, we conclude that
\begin{itemize}
\item[1.] Our proposed approach substantially increases the power of the detection when the employed eigenfunctions of $LC_X(t,s)$ cannot explain the change function.
\item[2.] When the employed eigenfunctions of $LC_X(t,s)$ can explain the change, the new detector still produces decent detection power. Thus there is no loss to apply the new detector. 
\end{itemize}
 In practice, it is tricky to know if the major eigenfunctions of $LC_X(t,s)$ can sufficiently explain $\delta(t)$, thus the change-aligned procedure is more reliable and likely to detect a true change-point.}

\subsection{Summary of Simulations}
The comparisons are summarized as follows.
\begin{itemize}
\item[1.] The type-I error of the proposed detector is well controlled around the nominal level under both $i.i.d.$~and dependent case. 
\item[2.] The performance of the proposed change-aligned detector is {\it robust} to the selection of $\rho$ and $\kappa$, and thus is not highly influenced by tuning parameters. 
\item[3.] The power of the proposed detector is obviously higher than that of the fully functional detector and the fPC-based detector, especially when the noise variation is high. The fPC-based detector typically gives the worst performance especially when the leading ordinary fPCs cannot explain the change. The fully functional approach, though performs better than the fPC-based approach, still gives suboptimal performance compared to the new approach especially when the noises become substantial. One explanation is that the fully functional approach incorporates the random noises into the detection procedure, and the nuisance effect of the random noises reduce the detection power.
It numerically demonstrates the necessities of careful selection of basis functions.
\end{itemize}

\section{Application to Annual Humidity Trajectories}
\label{s5}
\begin{figure}[ht]
\center
\includegraphics[scale=0.5]{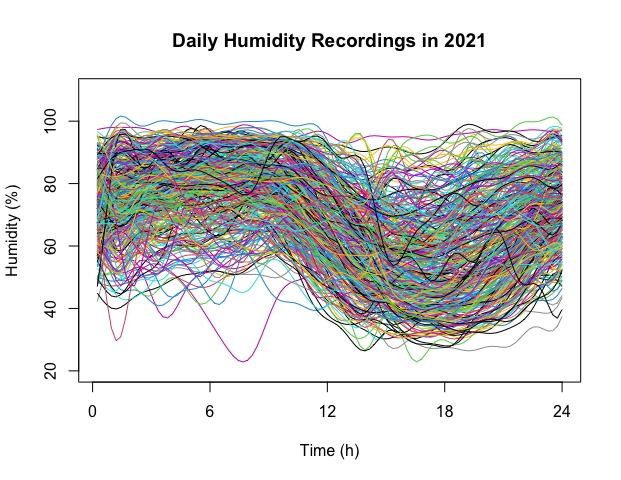}
\caption{Daily humidity curves in Basel-City, Switzerland.}
\label{humidity}
\end{figure}

In this section, the proposed approach is applied to daily humidity trajectories obtained in Basel-City, Switzerland in 2021. The raw data consist of $N=365$ daily measurements of humidity recordings (one observation per hour, 24 observations for each day) that are converted into functional objects by using 24 Fourier basis functions. The data can be downloaded at www.meteoblue.com. Figure \ref{humidity} displays the trajectories.
For comparison, the proposed detector and the other two competitors (the fPC-based detector and the fully functional detector) are applied to date the time of the structural breaks. 

\subsection{Dynamic Segmentation}
To attenuate the violation of at most one change-point assumption (AMOC), we first segment the entire sequence into multiple disjoint blocks. The segmentation approach employed here is motivated by the dynamic segmentation approach (see \cite{chiou2019identifying}) and is adapted for our own purpose, which is described below.

First we segment the whole functional sequences into 10 equal-length blocks 
$$\{[\theta^{(0)}_r,\theta^{(0)}_{r+1})\colon r=1,\ldots,10\},$$ where $\theta^{(0)}_1=1$ and $\theta^{(0)}_{11}=365$. Then recursively update the segment points as follows.

Given a subinterval $[\theta_r,\theta_{r+1})$ of $[1,365]$ and any $\theta$ in the subinterval, the sample covariance is calculated as follows 
\begin{align*}
S_{[\theta_r,\theta_{r+1})}^{(\theta)}(t,s)=\frac{1}{\lfloor{N\theta_{r+1}}\rfloor-\lfloor{N\theta_r}\rfloor}\sum_{n=\lfloor{N\theta_r}\rfloor}^{\lfloor{N\theta_{r+1}}\rfloor}\{X_n(t)-\bar{X}^{(\theta)}_n(t)\}\{X_n(s)-\bar{X}^{(\theta)}_n(s)\}\,,
\end{align*}
where
\begin{equation*}
\bar{X}^{(\theta)}_n(t)=\left\{
\begin{aligned}
&\frac{1}{\lfloor{N\theta}\rfloor-\lfloor{N\theta_r}\rfloor}\sum_{n=\lfloor{N\theta_r}\rfloor}^{\lfloor{N\theta}\rfloor}X_n,\ n\in [\lfloor N\theta_r\rfloor,\lfloor N\theta\rfloor]\,,\\
&\frac{1}{\lfloor{N\theta_{r+1}}\rfloor-\lfloor{N\theta}\rfloor}\sum_{n=\lfloor{N\theta}\rfloor+1}^{\lfloor{N\theta_{r+1}}\rfloor}X_n,\ n\in (\lfloor N\theta\rfloor,\lfloor N\theta_{r+1}\rfloor]\,.
\end{aligned} 
\right.
\end{equation*}
Suppose in the $i$-th interation, the segmentation points are $\{[\theta^{(i)}_r,\theta^{(i)}_{r+1})\colon r=1,\ldots,10\},$ where $\theta^{(i)}_1=1$ and $\theta^{(i)}_{11}=365$ for all iteration $i$. 
For each $r>1$, find the $\theta\in[\theta^{(i+1)}_{r-1},\theta^{(i)}_{r+1})$ that minimizes $\|S_{[\theta^{(i+1)}_{r-1},\theta^{(i)}_{r+1})}^{(\theta)}\|_\mathcal{S}$, which is set as ${\theta}^{(i+1)}_{r}$. The iteration stops when $\max_{1\le r\le 10}\abs{\theta^{(i+1)}_r-\theta^{(i)}_r}<1/N$. The final segmentation points are denoted by $\{\tilde{\theta}_r,r=2,\ldots,10\}$, and $\tilde{\theta}_1=1$ and $\tilde{\theta}_{11}=365$.

\begin{figure}[ht]
\center
\includegraphics[scale=0.5]{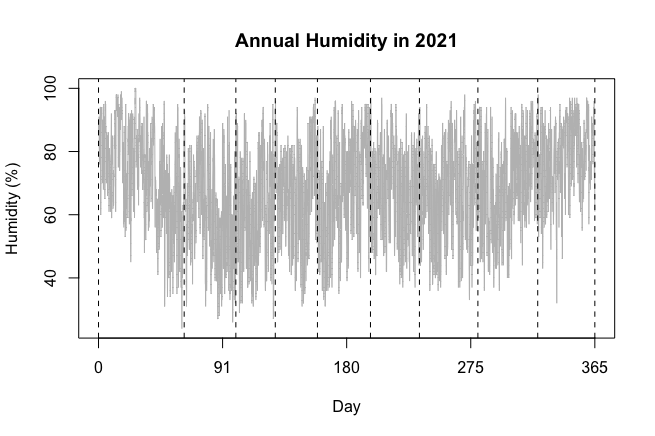}
\caption{Initial segmentation.}
\label{seg}
\end{figure}

{\it Our proposal} is that the whole sequence $[1,N]$ is segmented by $\{(\tilde{\theta}_r+\tilde{\theta}_{r+1})/2\colon r\ge2\}$. 
Note that, in \cite{chiou2019identifying}, $\{\tilde{\theta}_r,r=2,\ldots,10\}$ are considered as change-point candidates. Each candidate will be tested under the AMOC assumption, and the statistically nonsignificant ones are removed. Here we divide the sequence $[1,N]$ disjointly so that each segment contains one such candidate. The initial segmentation of $[1,365]$ is displayed in Figure \ref{seg}.

\subsection{Backward Elimination}
For each segment, we apply the three detectors considered in the simulation to detect and date the change-point under the AMOC assumption. If there is no change point detected in the subinterval $[{\theta}_r,{\theta}_{r+1}]$, then remove ${\theta}_{r+1}$ and test the change-point in the longer subinterval $[{\theta}_r,{\theta}_{r+2}]$. The elimination procedure stops till no segmentation point is removed. 

\begin{figure}[H]
\center
\includegraphics[scale=0.5]{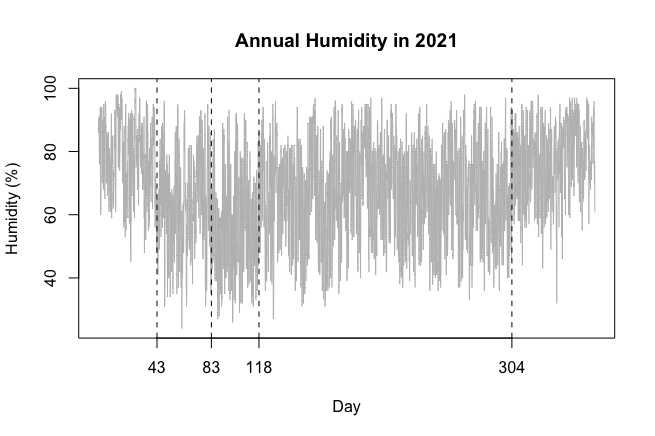}
\caption{Detected days of change-points.}
\label{final}
\end{figure}

\begin{figure}[H]
\center
\includegraphics[scale=0.5]{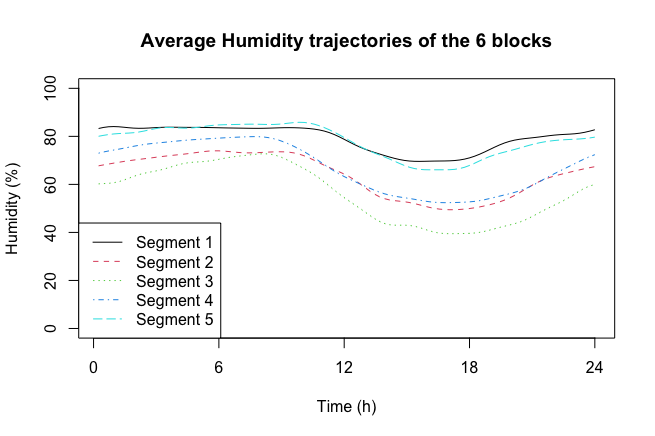}
\caption{Average Humidity Trajectories of each segment.}
\label{mean}
\end{figure}

Here, $\ell=3$, and $\rho=N^{0.4}$.
Both our approach and the fully functional approach detect 4 change-points at significance level $0.05$, which are displayed in Figure \ref{final}, while the fPC-based approach detect two change-points only, say, the 43th and 304th day of the year. The mean functions of the 5 segments are displayed in Figure \ref{mean}.

In this application, although the proposed detector and the fully functional detector work similarly, there are cases when our approach is superior to the fully functional approach. There is evidence to believe that the developed procedure offers a more reliable method to detect change points in functional means. 

\section{Conclusions}
\label{s6}
In this paper, a new change-aligned detector is introduced to detect and date the structural breaks in mean function of weakly dependent functional data. This detector has several advantages compared to the existing representative approaches including the fPC-based detector and the fully functional detector. 
Specifically, the fPC-based approach does not work while the employed fPCs fail to explain the structural breaks, and the fully functional approach essentially selects all basis functions that span the functional space, and thus suffers more from the nuisance effect of the irrelevant basis functions than the developed change-aligned procedure. The proposed detector relies on the carefully selected basis functions that are informative to the change in mean, making it more reliable to detect the change while controlling the type-I error close to the nominal level. In the simulation study, it is shown that the proposed detector performs better than the fully functional and fPC-based detectors, especially when the functions are contaminated by random errors or the leading fPCs cannot explain the change of mean.

\end{document}